\def   \ni {\noindent}

\def   \ssk {\vskip  5truept}

\def   \bsk {\vskip 15truept}

\def   \newline {\hfil\break}

\def\etal{ et al.~}

\def\KN{\scriptscriptstyle{KN}}

\def\M{\scriptscriptstyle{M}}

\def\etal{ et al.~}

\newcommand{\siml} {\lower.5ex\hbox{$\; \buildrel < \over \sim \;$}}
\newcommand{\simg} {\lower.5ex\hbox{$\; \buildrel > \over \sim \;$}}

\documentstyle[epsfig]{article}
\begin{document}

\hsize 5truein
\vsize 8truein
\font\abstract=cmr8
\font\keywords=cmr8
\font\caption=cmr8
\font\references=cmr8
\font\text=cmr10
\font\affiliation=cmssi10
\font\author=cmss10
\font\mc=cmss8
\font\title=cmssbx10 scaled\magstep2
\font\alcit=cmti7 scaled\magstephalf
\font\alcin=cmr6 
\font\ita=cmti8
\font\mma=cmr8
\def\ref{\par\noindent\hangindent 15pt}
\null


\title{\ni Probing the Emission Sites of GRBs}

\bsk \bsk
\author{\ni Hara Papathanassiou}                                                       
\bsk
\affiliation{ SISSA, Via Beirut n 2-4, 34013 Trieste, ITALY
}                                                
\bsk
\baselineskip = 12pt

\abstract{ABSTRACT \ni
I present a computationally efficient way to account for synchrotron and its
inverse Compton scattered emission along with  the resulting radiative losses, in a self-consistent way, for a relativistic distribution of electrons continuously injected in an expanding region. This approach permits  the exploration of the wide parameter space of the physical processes in a shock. I apply it to the picture of  ``internal shocks'' for
 the description of Gamma Ray Bursts. I use the properties of the time-integrated spectra as defined mainly by observations by BATSE, EGRET, and Ginga to impose constrains on the parameter space. I discuss the corresponding shock properties 
and one of the ways in which INTEGRAL can provide further insight.
}                                                    
\bsk
\baselineskip = 12pt
\keywords{\ni KEYWORDS: gamma rays: bursts ---gamma rays: theory ---radiation mechanisms: non-thermal 
}               

\bsk
\baselineskip = 12pt


\text{\ni 1. INTRODUCTION
\ssk
\ni     
A large database of Gamma Ray Bursts (GRB) has be accumulated by BATSE, 
over the 7.5 years of its operation.
The recent detections and follow-up observations of fading and softening counterparts by the BeppoSax satellite have furnished concrete proof of both the
{\it extragalactic origin} (e.g., Kulkarni \etal 1998) and {\it relativistic 
expansion} (Frail \etal 1998) of the sources. 
The BATSE database reveals a  rich morphology of   lightcurves as well as spectra. Most of the lightcurves are complex and highly structured pointing to an underlying
 {\it ``central engine''}. 
 The spectra are wide, non-thermal, and they usually exhibit one  spectral break, the position and slopes away from
 which have considerable scatter. The spectral shapes suggest {\it shock acceleration} and {\it non-thermal radiation processes}.

Based on these properties (proven or safely inferred), I examine the time integrated spectra resulting 
from synchrotron ({\it Sy}) and inverse Compton scattering ({\it IC}) from $e^-$ continuously injected in a propagating shock. The cooling of the assumed
 injected power law is calculated self-consistently. I develop a computationally 
efficient scheme that is appropriate for parameter searches in the wide dynamic range 
of the shock properties. I use the collective  spectral properties of GRBs as
established by the observations of BATSE, EGRET, and Ginga, for the purpose of 
constraining the physical parameters of the shocks.


Fitting of BATSE time integrated spectra (Band \etal 1993) by a broken power law
gives 
 $\alpha \in [-1.5, 2]$ and $\beta \in [-2.5, -1.5]$, for the slopes,
 while the spectral break  shows a spread between 
below 100 keV and above a few MeV.
Moreover, spectra of COMPTEL (Hanlon \etal 1994) and EGRET (Catelli \etal 1997)
bursts are fitted by a single power law of $-1.6 \siml \beta \siml -3.6$ 
(seemingly  an extension of the  BATSE high energy portion) with no indication
of a high frequency cut-off. Nevertheless, searches in the 10's of GeV to 100's of
TeV range proved fruitless (Aglietta \etal 1996).

To the soft side of the BATSE window, early experiments established 
an {\it X-ray paucity}, i.e., the emission in the few
keV range lies below a few percent of that in the few 100's of keV. Recently, there has been growing evidence that this behavior is not shared by the full body of GRBs. BATSE finds an {X-ray excess} rather than a deficiency below 20~keV in 15\% of
the examined sample (Preece \etal 1996) and so does Ginga in about 40\% of its sample (Strohmayer \etal 1997).

\bsk
\ni 2. THE PHYSICAL MODEL
\ssk
\ni 

I consider a  distribution of relativistic $e^-$ injected at a constant
number density (i.e., per volume rate) in a volume that is increasing at constant
speed. This could be a region in a flow that is being shocked by a sequence of
similar shocks, being energized continuously at its base and growing due to the
propagation of the first shock front. For the energy distribution of the $e^-$, I assume that a hard power law tail develops above $\gamma_{m,o}$. The power law index $p$ and the peak of the distribution $\gamma_{m,o}$ are free parameters.
For an injection  duration of $t_o$ and total number density $n_e$, the energy distribution of the rate is 
\begin{equation}
\dot{n}_{\gamma} \sim \frac{n_e}{t_o \gamma_{m,o}} 
\left\{ \begin{array}{l}
\frac{\gamma^2 -1}{\gamma_{m,o}^2 -1} e^{1-\gamma/\gamma_{m,o}}, \;\;\gamma < \gamma_{m,o} 
\\
 \left(\frac{\gamma}{\gamma_{m,o}}\right)^{-p}, \;\;\gamma_{m,o}\le \gamma \le \gamma_{{\M},o},
\label{eq:el_sp}
\end{array}
\right.
\end{equation}
where 
the hard end of the distribution $\gamma_{\M,o}$ is determined at every timestep by the size of the emitting region. All quantities are in the frame co-moving with the flow.

The distribution  cools under {\it Sy} and the {\it IC} scattering of 
it.
Fitting the resulting time integrated spectrum, one may express all its
 characteristic frequencies in terms of the corresponding  energies 
of the $e^-$ distribution, the magnetic field $B$, and  $p$.
Therefore, given the shape of the $e^-$ distribution at every instant that
follows from the injection prescription and cooling in the global magnetic and
radiation fields, one can describe parametrically the {\it Sy} component of the spectrum. The same can be done with the {\it IC} component, assuming  the spectral shape results from scattering of  monoenergetic radiation with the mean frequency of {\it Sy}.

For the  description of the time integrated spectra, the characteristic energies of the $e^-$ distribution are sufficient. In order to avoid the expensive computation of the evolution of the $e^-$ distribution at every timestep, with self-consistent account of the {\it IC} emission, along with the spectral distribution of the emission, the following procedure is adopted:
At every timestep, the evolution of the edges $\gamma_m$ and $\gamma_{\M}$ 
of the $e^-$ distribution is calculated given the total number density of $e^-$ present in the region and the radiation energy density and mean 
frequency of the synchrotron radiation produced at all previous timesteps. 
Cooling is evaluated in two limiting cases, accounting for {\it IC} scattering in the Thomson and the extreme Klein-Nishina (KN) regime, with an adaptive timestep to insure smooth transition.
The power law develops a knee which, at the end of the injection (assumed to last as
long as the emission), is at $max\{ \gamma_{m,o}, \gamma_{\M,t_o}\}$.
Apart from $\langle\,\gamma^2 \rangle\,_{t_o}$, the value $\gamma_{\KN}$ is 
computed were the distribution may flatten (this happens if losses in the extreme KN regime outweigh those of {\it Sy}).
From the evolution of the edges of the $e^-$ distribution, the total power emitted is also obtained;  the relative strength of the {\it Sy} and {\it IC} components is weighted by the cooling timescales of $\gamma_{m,o}$.

Tying the described physical system to the proposed picture of internal shocks
for GRB (M\'esz\'aros  \&  Rees 1994), I express the needed quantities in terms of the physical properties of the source, with appropriate parameterization. Internal shocks develop when a flow with an intrinsic variability on $t_{var}$ dissipates its
fluctuations. The flow is further characterized by a total energy $E_o$,
 solid angle $\Delta \Omega$, mass injection rate $\dot{M}_o$, and duration of the 
activity $t_w$ (which also determine the flow's bulk Lorenz factor $\Gamma \siml \eta \simeq E_o/\dot{M}_o t_w c^2$).
Such internal shocks develop at $r_d \simeq \Gamma^2 c t_{var}$ from the center of the activity. These determine the energy and mass densities at the dissipation
site to which the properties that are relevant to the emission processes are linked.
Consequently, the magnetic field energy density is expressed as a fraction $\lambda$ of the internal energy, the number density of the injected $e^-$ as a fraction $\zeta$ of the $p$ energy (assuming equal numbers of $p$ and $e^-$ in the flow), and the mean energy of the injected $e^-$ distribution by $\kappa$. The efficiency of passing energy from the main inertia carriers to $e^-$ during acceleration, is  $\varepsilon_{pe} = \zeta \kappa m_e/m_p$ with $\kappa \le \zeta m_p/m_e$. Also, in this picture, $t_o = \Gamma t_{var}$.

\bsk
\ni 3. PARAMETER SEARCH
\ssk
\ni

A large fraction of the time-integrated spectra shows a break in the BATSE window.
Furthermore, the self absorption frequency is scarcely met in this range. Fixing
the $e^-$ spectral index to $p=3$, close to the full range of fitted 
spectral slopes is obtained, depending on the characteristic $e^-$ energy that
falls in the BATSE window.
The criteria that I adopt to search for parameters that satisfy the constrains are set mainly by the data collected by BATSE, EGRET and Ginga -and will be referred to as BEG. These are:
1) The burst is ``detectable'' by BATSE, i.e.,
$(\nu {\cal{F}}_{\nu})_{300\rm{keV}} \ge 10^{-8} \rm{erg/cm^2}$.
2) Losses in the KN regime should not affect the shape of the $e^-$ distribution in the range  probed by BATSE.
3) One of the spectral breaks of either the {\it Sy} or the {\it IC} component that
correspond to $\gamma_{m,t_o}$, $\gamma_{m,o}$, or $\gamma_{knee}$, or, the frequency below which the {\it Sy} turns over to the Rayleigh-Jeans slope, or its
{\it IC} upscattered counterpart falls in the BATSE window.
4)
The photon number index in the EGRET range, up to 300 MeV is $\beta_{300\rm{GeV}} \siml -1.6$.
5)
The burst is X-ray deficient: $(\nu {\cal{F}}_{\nu})_{2\rm{keV}} \le 5\% (\nu {\cal{F}}_{\nu})_{300\rm{keV}}$.
6)
The burst is X-ray rich: $5\% (\nu {\cal{F}}_{\nu})_{300\rm{keV}} \le
(\nu {\cal{F}}_{\nu})_{2\rm{keV}} \le 2 (\nu {\cal{F}}_{\nu})_{300\rm{keV}}$.

I adopt $E_o =10^{51}$~erg, $\Delta \Omega = 0.12$~rad and $100 \le \Gamma \le 300$, in accordance with estimates implied by afterglow observations.
Searches in the shock  parameter space (i.e., $\kappa, 
\lambda$, and $\zeta$) are conducted with the following combinations of the BEG criteria:
1; 1, 2, and 3; 1, 2, 3, and 4; 1, 2, 3, and 5; 1, 2, 3, and 6.

\bsk
\ni 4. RESULTS AND DISCUSSION
\ssk
\ni 

Even the most restrictive of the parameter regions allow for a wide range of spectral shapes. Only for the highest possible magnetic field values does the {\it Sy} component lie in the BATSE window and, in most of these
cases,  copious photon production results in  a prominent component (pair absorbed, in cases) in the EGRET window.
 This is inconsistent with observations.
Upon closer inspection of the resulting spectra, many parameters can be further excluded based on the broader spectra features. 
Requiring the thermal slope to be way below the BATSE window leaves very little
allowed parameter space indeed. Hence, I tentatively conclude that the transition to the
optically thick part of the spectrum happens at a frequency fairly close to the
low edge of BATSE. This is in agreement with the suggestion of the Ginga spectra
analysis of a second spectral break around 5 keV (Strohmayer \etal 1997).

The present analysis suggests that {\it IC} is most often the BATSE component.
A rather narrow range of the $e^-$ distribution parameters corresponding to a low efficiency for transferring of energy from $p$ to $e^-$ ($0.1\% \siml \varepsilon_{pe} \siml 1\%$), combined with a wide range in the fraction of the magnetic field energy density ($10^{-8}\siml 
\lambda \siml 10^{-2}$) can reproduce the full range of the documented properties.
With {\it IC} being in the  BATSE  range,  higher $B$ field values lead to a more pronounced and narrowly spaced 
{\it Sy} component thus resulting in an X-ray excess case, while weaker field suppresses {\it Sy} emission exposing the steep (thermal slope) part of the {\it IC} component therefore listing the case as an X-ray deficient one.

Although this is not a comprehensive parameter search, these observations should hold roughly unchanged in general too. There is no strong dependence of the allowed parameter space on $t_{var}$, $\Gamma$ is tightly constrained in a
narrow range around 300, and the shock properties should not depend on the central engine operation lifetime or total energy (apart from affecting detectability).

One of the suggestions of this  analysis is that  the transition to the optically thick portion
of the time integrated spectra, happens close to the low end of the BATSE window
(some times inside it). 
Extending the simultaneous spectral coverage down to the keV range 
(which is in the INTEGRAL capabilities)  will help answer this question thus 
providing valuable insight into the conditions of the associated shocks.


\bsk
\baselineskip = 12pt


{\references \ni REFERENCES
\ssk}
%
\ref Aglietta \etal 1996, ApJ, 469, 305
\ref Band \etal 1993, ApJ, 413, 281
\ref Catelli \etal 1997  in {\it Proc. of the 4th Huntsville Symposium on GRBs}
\ref Frail \etal 1997, Nature, 389, 261
\ref Hanlon \etal 1994, A\&A, 285, 161
\ref Kulkarni \etal 1998, Nature, 393, 35 \label{z=3.5} 
\ref M\'esz\'aros  \&  Rees 1994, MNRAS, 269, L41
\ref Preece et al. ApJ, 473, 310
\ref Strohmayer \etal 1997, preprint (astro-ph/9712332)


\end{document}